# Electron-hole Hybridization in Bilayer Graphene

Siqi Wang[1 †], Xianqing Lin[2, 3 †], Mervin Zhao[1], Changjian Zhang[4], Sui Yang[1], Yuan Wang[1], Kenji Watanabe[5], Takashi Taniguchi[5], James Hone[4], David Tomanek[2], Xiang Zhang[1 *]

**Affiliations**

[1]*NSF Nanoscale Science and Engineering Center (NSEC), 3112 Etcheverry Hall, University of California, Berkeley, California 94720, USA.*

[2]*Physics and Astronomy Department, Michigan State University, East Lansing, Michigan 48824, USA*

[3]*College of Science, Zhejiang University of Technology, Hangzhou 310023, China*

[4]*Department of Mechanical Engineering, Columbia University, New York, NY, 10027, USA*

[5]*National Institute for Materials Science, 1-1 Namiki, Tsukuba, Japan*

[†] *These authors contribute equally*

[*] *Correspondence to xzhang@me.berkeley.edu*

**Band structure determines the motion of electrons in a solid, giving rise to exotic phenomena when properly engineered. Drawing an analogy between electrons and photons, artificially designed optical lattices indicate the possibility of a similar band modulation effect in graphene systems. Yet due to the fermionic nature of electrons, modulated electronic systems promise far richer categories of behaviors than those found in optical lattices. Here, we uncovered a strong modulation of electronic states in bilayer graphene subject to periodic potentials. We observed for the first time the hybridization of electron and hole sub-bands, resulting in local band gaps at both primary and secondary charge neutrality points. Such hybridization leads to the formation of flat bands, enabling the study of correlated effects in graphene systems. This work may also offer a viable platform to form and continuously tune Majorana zero modes, which is important to the realization of topological quantum computation.**

**Main text**

Behavior of ballistic electrons in a uniform material resembles that of photons to a high degree [1-7, 35]. For example, electrons follow straight trajectories when considered as particles [16-18], while interference effects, such as the Aharonov-Bohm [19-21] and Fabry-Perot effects [22], are caused by their wave nature. Due to the conservation of the transverse momentum and the Fermi energy, electron propagation at the boundary of two regions with different carrier densities is subject to reflection and refraction in a way similar to optical rays crossing the boundary of two materials with different refractive indices [23]. Unlike photons, electrons in an atomically thin material can be efficiently manipulated by an artificially designed and applied potential profile that controls the spatial carrier density profile [24-26], resulting in graphene electron optics. This opened the way to realizing scenarios that are typically difficult to achieve by traditional optics, such as negative refractive index and non-linear dispersion, in solid-state systems like graphene [27, 28].

So far, research in graphene electron optics has focused on the single-interface effects, including the Veselago lens [23, 27, 28] and whispering gallery [29]. However, modern optics has utilized the concept of superlattices (SL) in photonic crystals, to achieve new optical phenomena. For example, the modulation of photonic states in one dimension by periodicity dramatically alters the photonic dispersion relation, which leads to commonly used dielectric mirrors, or the distributed Bragg reflector [30]. Moreover, in-plane photonic SLs have been utilized to demonstrate novel physics, such as topological electromagnetic states [36]. However, there are few experimental studies of SL effects in electronic systems, majorly in periodic quantum well structure [31]. Yet it limits the electrostatic modulation only along its growth axis. In this work, by employing a bipolar SL modulation in high-quality BLG, which exhibits a touching-hyperbolic dispersion that can be effectively modified electrostatically, we discovered a strong band modulation effect. For the first time, we demonstrate hybridization of electron and hole sub-bands causing emergence of local band gaps of several milli-electron-Volts around both the primary and secondary charge neutrality points. Such a hybridization leads to the formation of flat bands around local gaps, providing a platform for the study of strongly correlated effects in graphene systems [2-4]. Since electronic states around local gaps are superpositions of electron and hole states, our experimental configuration may allow to observe and continuously tune Majorana zero modes, which may find an application in topological quantum computation due to their non-abelian statistics [11-15].

Similar to a photonic crystal [9, 10], artificial periodicity in a 2D material system induces band folding in the underlying dispersion of the pristine lattice [2, 32]. In the case of BLG, the original hyperbolic dispersion is folded into the reduced Brillouin zone, forming sub-bands as shown in Fig. 1a,b. Moreover, in an electronic system, changes in the potential profile shift the electron sub-bands and hole sub-bands relative to the Fermi level, causing charge doping [33]. Depending on the sign of the potential profile, in general, there are two modulation regimes, namely monopolar modulation (Fig. 1c left), where carriers of the same type are doped into the system, and bipolar modulation (Fig. 1c right), where both types of carriers are doped in different spatial regions. While the monopolar modulation shifts electron and hole sub-bands similar to the gating effect of a uniformly gated BLG, we show that bipolar modulation causes intersection of electron and hole sub-bands due to the simultaneous appearance of electron and hole states around the Fermi surface. In this scenario, the dispersion of BLG is significantly modified. As a consequence, confirmed by our charge transport measurements and calculations, local band gaps open at the crossing of electron and hole sub-bands resulting in sub-band hybridization in BLG.

The encapsulation of 2D materials between hexagonal boron nitride (h-BN) can greatly improve device performance and protect the material from unwanted contamination [8]. Similarly, we encapsulated BLG between h-BN flakes using the van der Waals dry transfer technique and the periodical SL was create electrostatically (Fig. 1d-f). The featureless back gate homogeneously doped the channel, whereas the periodic top gate independently induced density modulation in the BLG. In order to align the Fermi level at different positions with the ground, the electrons in BLG were subjected to an electrostatic potential U(x), which fulfills $-eU(x) + E(k_F(x)) = 0$. The precise potential profile depends on the thickness of h-BN flakes, the top-gate voltage $V_{tg}$, and the back-gate voltage $V_{bg}$. We obtained a potential profile representative of our device parameters using a self-consistent calculation that takes the quantum nature of the BLG channel into account (Fig. 2c).

The electronic modulation can be directly measured by electron transport in the device using four terminal lock-in technique at a temperature of 1.8 K. Transport measurements from a device with a SL period of 120 nm are shown in Fig.2a, where the measured resistance is plotted as a function of $V_{tg}$ and $V_{bg}$. Applying a $V_{tg}$ "turns on" the SL potential, and two patterns can be observed. The first feature to note is the appearance of a bright vertical line at $V_{bg}=0$, corresponding to the primary charge neutral point (CNP). Since the graphitic back gate extends across both the channel and contact areas, whereas the top-gate covers only the channel region (Fig. 1d), the position of the bright vertical line is independent of top-gate tuning.

More interestingly, two groups of clear patterns were observed to fan out diagonally from the CNP. This is very different from transport results obtained using a conventional BLG device with a featureless top gate and back gate, where we would expect only one diagonal line corresponding to the tuning of central CNP. In our device with a patterned top gate, we identified eight regions (I – VIII in Fig. 2a) around the main CNP and interpreted them as eight doping regimes induced by the two gates. In the typical monopolar modulation regime, region I (V), two gates electrostatically doped electrons (holes) into BLG resulting in low resistance. A similar situation occurred at weak SL potentials (region II or VI), where the resistance was not significantly changed by the top-gate because it did not affect carriers of the opposite type. As we increased the SL such that $V_{tg}$ and $V_{bg}$ were comparable in strength (region III or VII), the bipolar doping regime, associated with region II to IV (VI to VIII), was achieved. The contact and channels became electron (hole) doped from the back-gate, whereas the top-gate selectively doped holes (electrons) electrostatically. Electron and hole states appeared at the Fermi level along the SL direction. Therefore, the effective potential $U(x)$ oscillated between positive and negative values as a function of the position x. These two regions were found where the strongest SL modulation was expected, resulting in the clear woven pattern in the fan area.

To understand the origin of the woven pattern, we calculated the transport properties of the device with the effective Hamiltonian $H = H_0(k_x, k_y) + U(x)$, where $H_0(k_x, k_y)$ is the effective Hamiltonian of BLG [34]. The electrostatic potential $U(x)$, shown in Fig. 2c, we obtained numerically in a self-consistent manner that takes the quantum nature of the BLG channel into account. Due to the very large size of 120 nm long unit cells, this nontrivial calculation utilized a previously unexplored numerical approach. Our results show that the original hyperbolic BLG band structure has been greatly modified by the SL potential (Fig. 2e). Spatial changes in the effective potential created an overlap between original electron-like and hole-like bands and caused them to intersect. Local gaps opened in the electronic spectrum around $k_y = 0$, with no states in the gap region at any value of $k_x$. Even though allowed electronic states still exist at large $|k_y|$, they do not contribute to transport in the device, since the bandwidth along $k_x$ rapidly shrinks to zero. When the Fermi level lies inside (outside) the newly created local gap, which can be achieved by tuning $V_{bg}$ and $V_{tg}$, the device has higher (lower) resistance.

In the experiment, at each negative $V_{bg}$, as the SL potential increased, the overlap between electron-like and hole-like bands increased gradually, which caused different local gaps to coincide with the Fermi level. It is this oscillatory nature of overlaps between the two bands that lies at the origin of the observed woven pattern. Indeed, the observed results were reproduced accurately in our numerical calculation, as seen in Fig. 2b. The mechanism of local gap formation is more directly revealed by analyzing the band character. As shown in Fig. 2e, states far away from local gaps were generally comprised of either electron or hole states, whereas states around local gaps appeared as a superposition. It is the introduction of the periodic potential that led to the intersection and hybridization of electron and hole sub-bands form new electronic states in the reduced Brillouin zone. The hybridization resulted in the appearance of local band gaps and mixed states.

We also noticed the appearance of a weak resistance modulation in regions I and V. Interestingly, as we increased $V_{bg}$, the weak pattern could be identified as part of a similar woven pattern centered at $V_{bg}= \pm 1.5$ V. In the full-range resistance map (Fig. 3a), we observed three sets of repeated patterns in total. We attributed the two new bright vertical lines to secondary CNPs of BLG, which were caused by the Moiré pattern forming between BLG and h-BN. According to the above analysis, the SL effect was imposed into the system via carrier density modulation. Similar gap-opening effects may also occur elsewhere, as long as the carrier modulation remains bipolar. Around the secondary CNPs, electrons and holes can be doped into the system as easily as near the primary CNP, which we have indeed achieved in the experiment.

On the other hand, since the potential in BLG is modulated only along one direction, we do not expect any change in transport properties normal to that direction. To confirm this reasoning, we fabricated another device with a different geometry, which allows to measure transport both along the direction of modulation and normal to that direction. Our results are shown in Fig. 3b, c. Whereas the resistance along the direction of the modulation (Fig. 3b) showed a similar woven pattern as that displayed in Fig. 2a, the resistance map along a direction normal to that (Fig. 3c) indicated merely a shift of the CNP under the combined influence of the top and the back gate.

Formation of local gaps in the spectrum was further confirmed by temperature-dependent measurements. Changes in the channel resistance at $V_{BG}$ = -0.55V as a function of $V_{TG}$ are shown in Fig. 4 in the temperature range from 1.8 K to 51.8 K. The oscillation in the woven pattern becomes flatter at higher temperatures, indicating that the effect of the SL potential decreases at higher temperatures due to thermal fluctuations, causing excitation of electrons across local band gaps at elevated temperatures. At high temperatures, the woven pattern in the resistance map washes out and changes to a normal high-resistance behavior of BLG transistors with a featureless dual gate. The woven pattern is completely suppressed beyond about 25 K, corresponding to a gap or activation energy of 2 meV, in quantitative agreement with the theoretical results. We consider this finding to provide independent support for the adequacy of the theoretical description of this complex system that matches well experimental results.

**Conclusion**

We have demonstrated strong modification of the BLG band structure by an artificial, electrostatically induced SL potential. Periodicity imposed by the SL led to band folding in the original band dispersion of BLG. In a regime of bipolar carrier density modulation, electron and hole states strongly hybridize and locally open band gaps around the Fermi level, which can persists up to 25 K. The bipolar modulation of BLG also leads to the observation of charge neutral electronic states composed of both electrons and holes, a signature of Majorana fermions and their quasiparticle bound states. The demonstrated electrostatically induced artificial SL potential profiles thus leads to a new venue to study novel electronic states Majorana physics. Owing to the large flexibility in the device design, the paradigm of modulating the electrostatic potential in the channel region may be generalized to higher dimensions and utilized to realize more exotic electronic phenomena.

**Methods**

The h-BN flakes were chosen with thickness ranging between 10 and 30 nm to ensure excellent gate insulation and modulation amplitude. The h-BN/BLG/h-BN stack was then transferred on few-layer graphite, which served as the conductive back gate in the measurement. To create our electrostatically defined potential, we used electron beam lithography to define a SL with a period of 120 nm and linewidth of 25 nm. Prior to the standard PMMA spin coating process, the stack was treated with mild oxygen plasma and HMDS to enhance the adhesion between the top h-BN surface and PMMA. Transport measurements were performed in Physical Property Measurement System from Quantum Design with a base temperature of 1.8K. Our four-probe measurements employed both low-frequency lock-in techniques (13.3 Hz frequency and AC excitation current of 10 to 100 nA) and DC measurement, both of which gave consistent results.

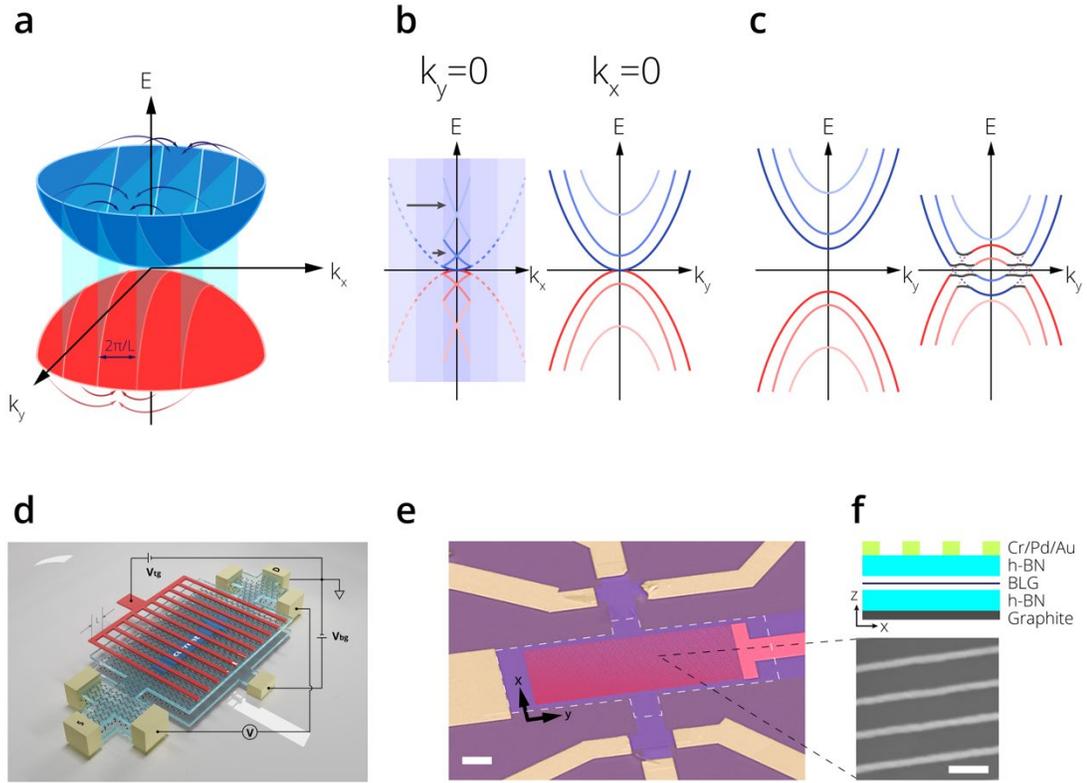

**Figure 1: Band modification and artificial one-dimensional superlattice in a BLG field-effect transistor with dual gates. a**, Schematic of band folding of BLG dispersion with an artificial periodicity L along the x direction, where the length of the reduced Brillouin zone along x becomes $2\pi/L$. **b**, While bands along $k_x$ are folded into the reduced Brillouin zone, electron and hole sub-bands extend along $k_y$. **c**, Two different modulation regimes. In the monopolar regime, the same type of carriers was doped into the system and the bands were shifted same as in the case of homogeneous dual gate doping (left). In the bipolar doping regime, electron and hole states appeared at the Fermi surface simultaneously and sub-bands overlap. Local band gaps appeared at the intersections. **d**, Schematic perspective view of the device. **e**, SEM image of a dual gate FET device (scale bar: 2 μm). BLG (blue) and the graphite bottom gate (marked by white dashed line) were connected via metal electrodes (yellow) via edge contacts. The metallic periodic top gate (red) covered the channel region. The scale bar is 2 μm. **f**, Schematic cross-sectional view of the device

and SEM image of the periodic top gate (scale bar: 100 nm). The period of the SL was 120 nm with a metal width of 25 nm. The top h-BN had a thickness of 25 nm and the bottom one had a thickness of 11 nm.

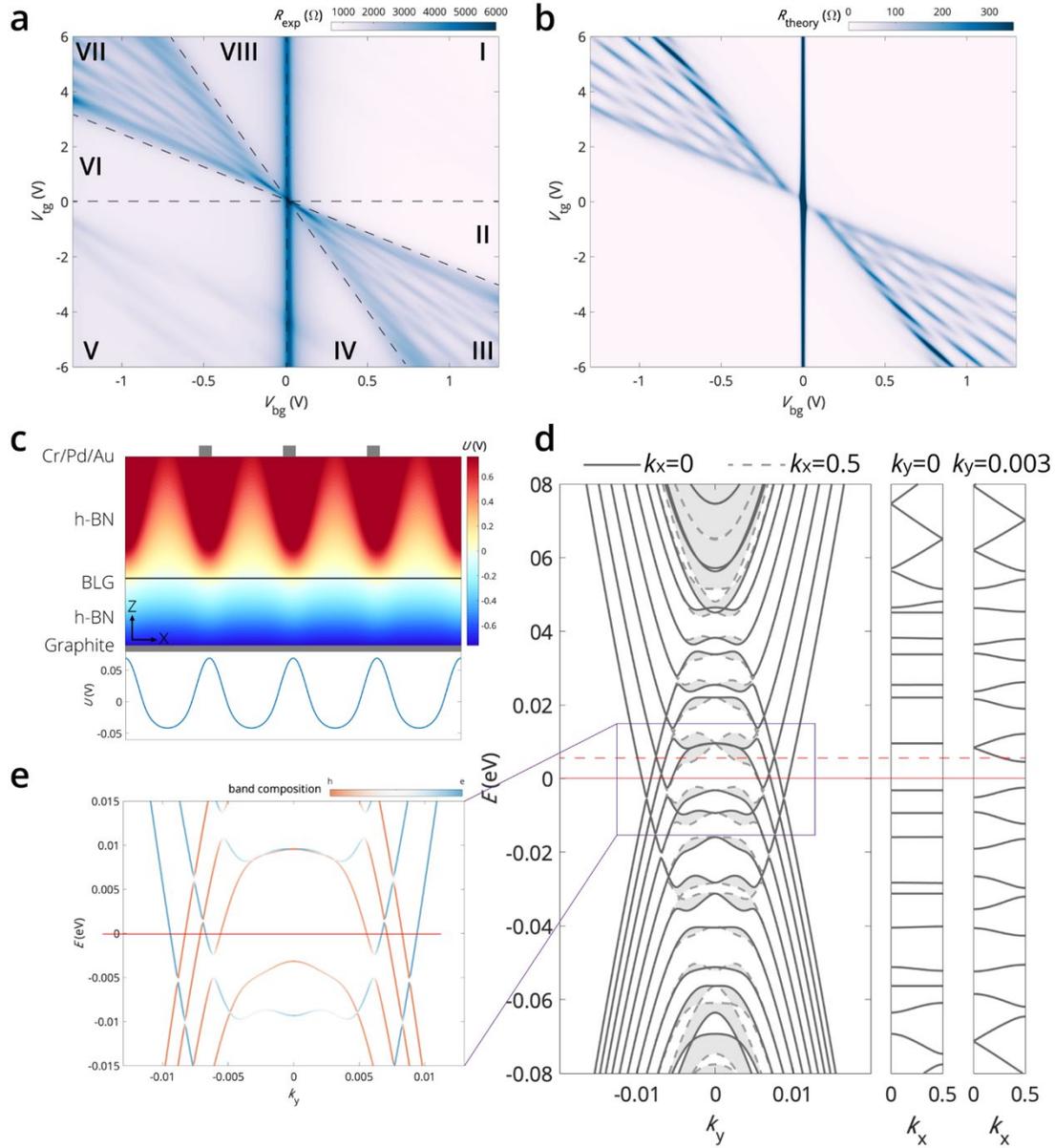

**Figure 2: Transport behavior in a BLG SL. a**, Color plot of the longitudinal resistance as a function of $V_{tg}$ and $V_{bg}$. The black dashed lines separate different doping regimes of top and bottom

gates. **b**, Calculated resistance color map as a function a $V_{tg}$ and $V_{bg}$. **c**, Simulated effective potential distribution at $V_{bg}$ = -0.11 V and $V_{tg}$ = 5.43 V across the device (above) and on BLG (bottom). **d**, SL electronic band structures at $V_{bg}$ = −1.11 V and $V_{tg}$ = 5.43 V. Dark grey solid lines and light grey dashed lines show bands along $k_y$ for $k_x$ = 0 and $k_x$ = 0.5 with $k_x$ and $k_y$ in fractional coordinates of the reciprocal lattice. The shaded regions represent the energy ranges of the bands along $k_x$ for each $k_y$. Red solid lines and red dashed lines denote the zero-energy levels and the neutrality-energy levels. **e**, Detailed band structure of the area marked by purple box in **d**. The SL potential has a period much larger than the period of the BLG lattice, which causes band folding in the $k_x$ direction. The original two-dimensional band structure was folded multiple times into the reduced Brillouin zone of the BLG SL, which manifested itself in the multi-band structure along the $k_y$ direction. The eigenstate composition is represented by the line color ranging from red (hole) to blue (electron). Mixed states appeared close to local gaps.

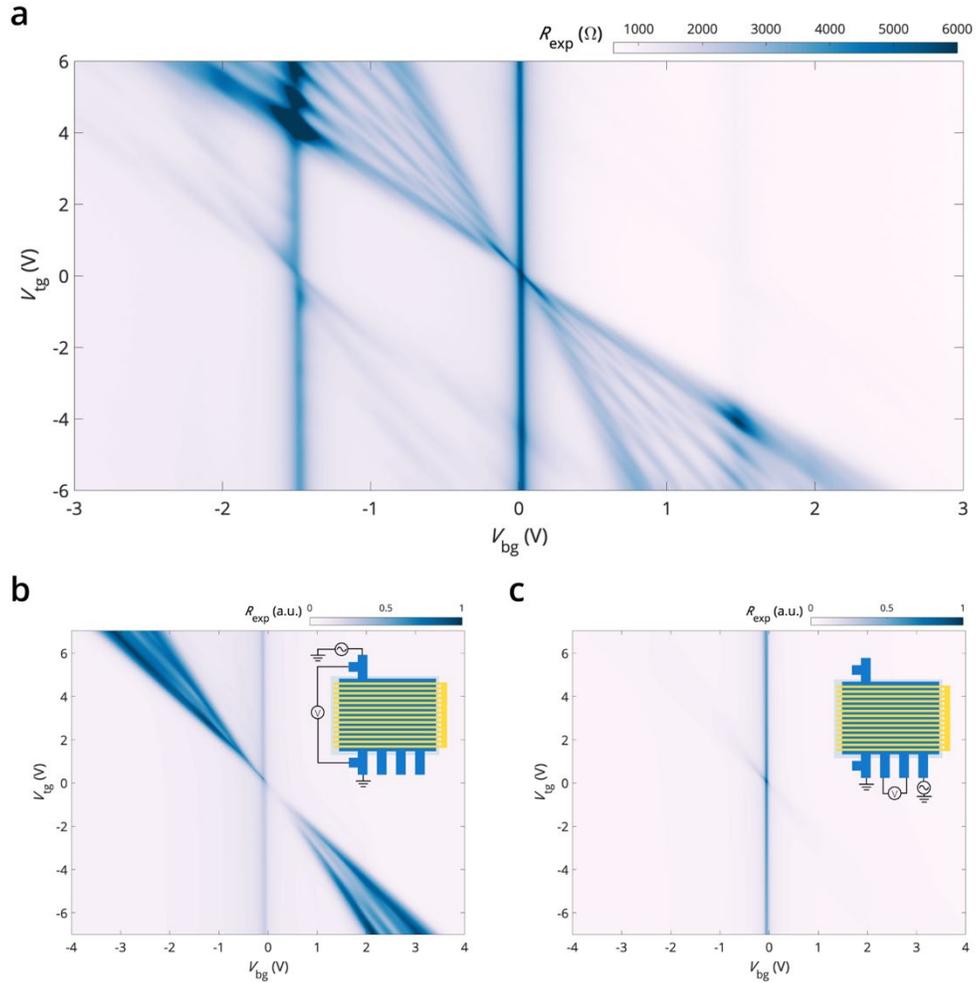

**Figure 3: Full range resistance mapping and the direction dependence of SL modulation effect.**

**a**, Full range resistance mapping showed three repeated sets of patterns originating from the main CNP at $(V_{bg}, V_{tg}) = (0, 0)$, and secondary CNPs at $(\pm 1.5\ V, 0)$. **b**, Resistance mapping with the current parallel to the periodicity showed a similar modulation pattern as in **a**. **c**, Resistance mapping with the current perpendicular to the periodicity only showed very weak modulation possibly due to imperfect geometry of the device or stray current. Insets in **b** and **c** are device and measurement schematics.

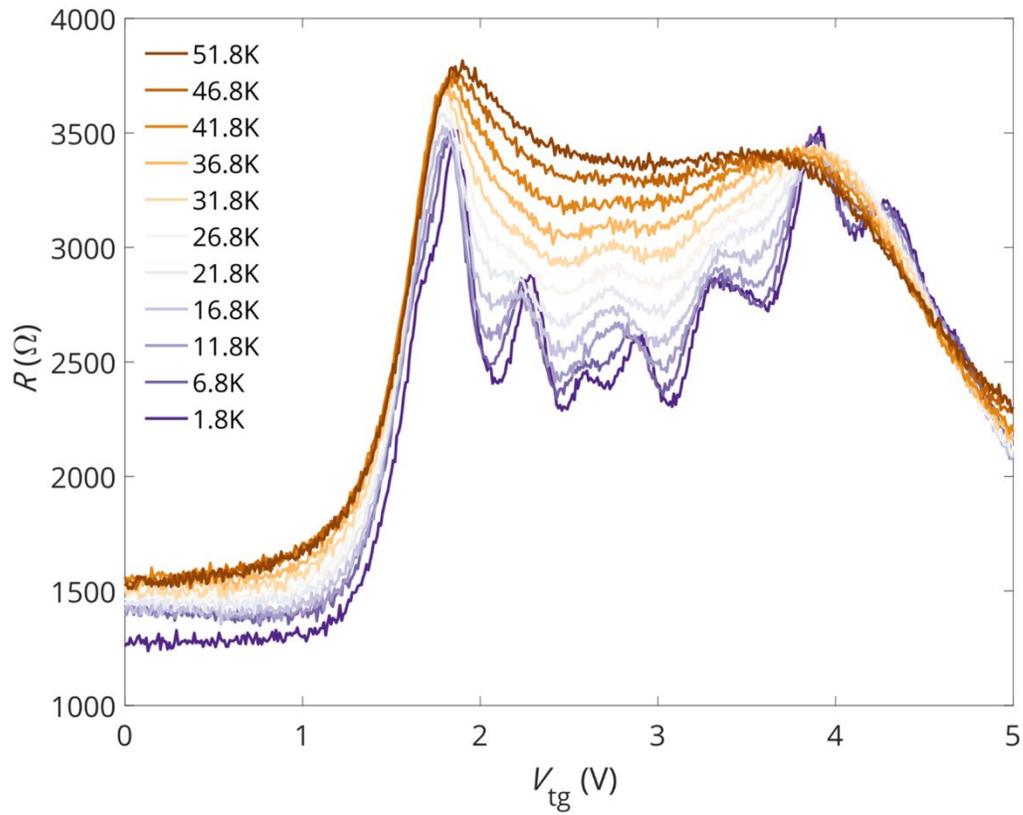

**Figure 4: Temperature dependence of BLG SL FET.** Perpendicular resistance of the device shown in Fig. 2a at $V_{bg}$ = -0.55 V with temperature ranging from 1.8 K to 51.8 K. The oscillation in the woven pattern became flatter at higher temperature and smears out around 25 K, corresponding to an activation energy of 2 meV.